\documentstyle[12pt,moriond]{article}
\begin{document}

\heading{Cosmic evolution and unified models for radio AGN}

\author{J.V. Wall $^{1}$, C.A. Jackson $^{2}$} {$^{1}$ Department of
Astrophysics, University of Oxford, Nuclear and Astrophysics Laboratory,
Keble Road, Oxford, UK OX1 3RH.} {$^{2}$ Mount Stromlo Observatory, The
Australian National University, P.O. Private Bag, Weston Creek, ACT,
Australia 2611.}

\begin{moriondabstract}

A new method of deriving the cosmological evolution of radio AGN is
described which is based on the premise of  unified models for quasars and
radio galaxies.

\end{moriondabstract}

The `starburst vs monster' debate takes place on several levels, for
example the relation between starburst and AGN nuclear
activity\cite{ter90} and between background contributions\cite{hec99}.  
The original version of the Madau diagram \cite{mad96} led several authors
in early 1997 to note the similarity between the cosmic histories of AGN
and star formation rate \cite{wal98,boy98}. The Madau diagram has evolved
\cite{ste99}, as has our view of how to determine space densities of AGN
\cite{wal97,jac99}. Here we briefly set out this new view of the cosmic
history of radio AGN, a history which must be correctly described if the
relation between AGN and star-formation activity is to be understood.

Previous analyses assumed two radio-AGN populations (e.g. \cite{dun90})
based on radio spectrum:  `flat-spectrum' (core-dominated, and
predominantly quasars) and `steep-spectrum' (extended emission, double
structures, and predominantly radio galaxies). In view of unification
models which posit that radio quasars and BL\,Lac objects are end-on
versions of radio galaxies, such a dichotomy makes no sense; and the
derivations of space densities have dubious interpretation.

\section{The paradigm}

Two anisotropic effects give rise to the unified view of quasars and radio
galaxies \cite{sch87,bar89}: relativistically-beamed twin jets feed the
double lobes of powerful radio sources \cite{bla78}, and the
black-hole/accretion disk system is shrouded in a dusty torus whose axis
is aligned with the radio axis \cite{ant93}. A radio galaxy with double
lobes is seen when the system is viewed side-on. As lines-of-sight
approach the axis, the torus opening reveals the light of the nuclear
black-hole/accretion-disk system which comes to dominate the galaxy light
to produce a quasi-stellar object. When lines-of-sight coincide closely
with the axis, Doppler enhancement of the relativistically-approaching
radio jet leads to its compact flat-spectrum radio emission dominating the
extended emission. Such `core-dominated' quasars show superluminal motions
in the jet structures as revealed by repeated VLBI observations (e.g.
\cite{pea87}).

Recognition of these two mechanisms has given rise to the two current
paradigms of radio-source unification, based on FRI and FRII radio
galaxies \cite{fan74} as the two parent populations. The FRI radio
galaxies show the two regions of highest surface brightness in radio
emission along the jets
feeding the double radio lobes; they are generally less powerful than the
FRIIs, and do not show strong optical/UV emission lines.  The
core-dominated counterparts are BL\,Lac objects \cite{bro83,mor91}.  The
powerful FRII galaxies show the brightest regions at the extremities of
the double lobes and have strong emission lines; their projected 
counterparts
are steep-spectrum quasars (at angles to the line-of-sight permitting a
view of the nucleus), and core-dominated quasars when the line-of-sight
coincides closely with the radio axis.

\begin{figure}[h]
\vspace{6.0cm}
\includegraphics{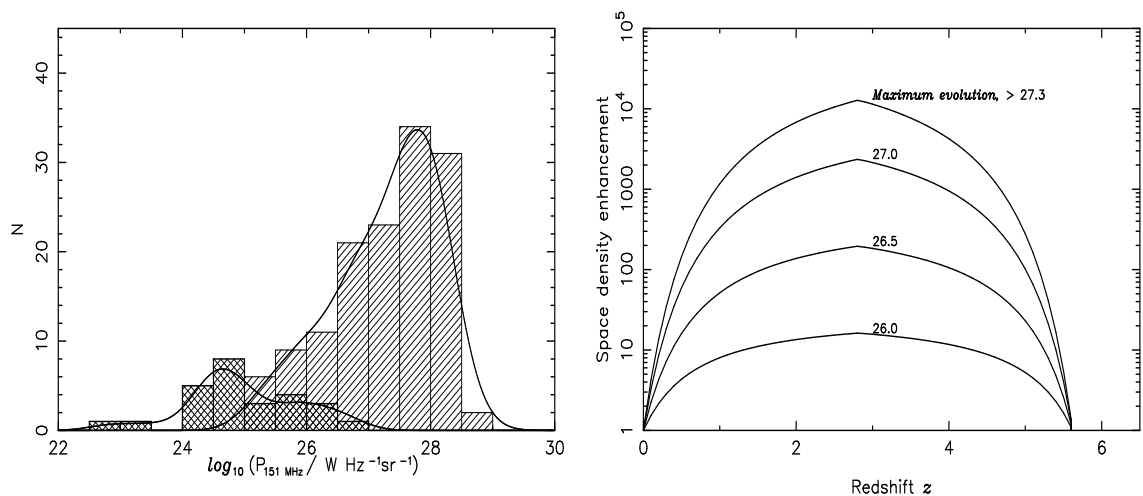}

\caption{Left: Luminosity distributions for the 26 FRI (cross-hatched) and
the 137 FRII (hatched) radio sources in 3CRR. Right: Space density
enhancements for the FRII parent population determined from optimized
model parameters over the range of log($P_{151 {\rm\ MHz}}$) shown.}

\label{lumdis}
\end{figure}

The first step in our analysis (described in detail in \cite{wal97,jac99})
is to estimate space densities as a function of epoch for the two {\it
isotropically-radiating parent populations}, the FRI and FRII radio
galaxies. As these objects dominate low-frequency radio surveys, we use
counts from the 3C and 6C (151 MHz) surveys and the
3CRR
luminosity distribution (\cite{lai83} and R. Laing, private communication;
see Figure~\ref{lumdis}) to derive space densities following the procedure
of Wall et al. \cite{wal80}. A parametric representation of
luminosity-dependent density evolution (Figure~\ref{lumdis}) is chosen to
mimic
the evolution found by Shaver et al. \cite{sha96}, and we determine the
best-fit parameters through a downhill simplex minimization process.

\begin{figure}[h]
\vspace{7cm}
\includegraphics{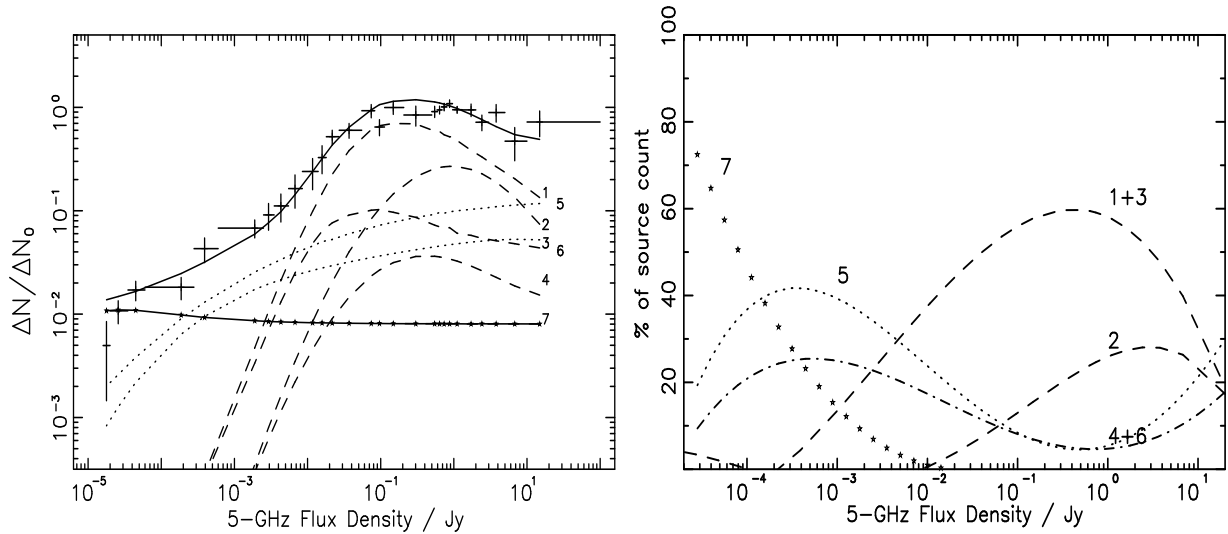}

\caption{Left: the 5-GHz source count in relative differential form. The
solid line shows the fit from optimization of the space density
description and beaming parameters. Populations~1 and~3 represent FRII
galaxies, populations~2 and~4 the relativistically-beamed quasars and
BL\,Lacs from these parents; population 5 the FRI parent population; and
population 6 the BL\,Lac objects, the beamed products from these. Line~7
represents the starburst-galaxy population which comes to dominate counts
below 1~mJy. Right: the proportions of these same populations making up
the total count at all flux levels. Note the rapid increase of starburst
objects below 1~mJy.}

\label{count}
\end{figure}

The second step was to `beam' these parent populations to determine the
contribution they make to the beamed flat-spectrum populations found in
higher-frequency ($\nu~\geq~1$ GHz) surveys. We adopt the simplest
possible beaming models, characterised for each of the two populations by
two parameters, a Lorentz factor describing the speed of ejection and the
ratio of the (rest-frame) fraction of beamed core emission to total
emission. Using Monte Carlo runs randomly-orienting the parent sources,
source-count predictions together with the proportion of beamed objects
involved can be made at all frequencies.  We use the minimization
procedure to determine the beaming parameters providing the best
prediction of the source counts at 5~GHz (Figure~\ref{count}).

\section{Results and discussion}

For the {\it parent populations}, the new analysis of space densities now
based on their completely-defined luminosity distributions finds three
essential features. (1) Powerful evolution is required for the FRII
population; in our parametric representation of evolution as $\exp(M\tau)$
with $\tau$ as look-back-time, $M_{\rm max}~=~10.9$ for the most
radio-luminous FRII galaxies.  (2) This density enhancement peaks
at~$z_c/2$ and tapers off to a redshift cutoff~$z_c$
(Figure~\ref{lumdis}). We obtain z$_c$~=~5.6; the fit with this value is
superior to the fit with no redshift cutoff at the 99.9\% level of
significance. (3) The FRI population shows little or no evolution ($M
\approx 0$), in agreement with the relative uniformity of space density
found for BL\,Lac objects (\cite{cav99} and references therein).

As for the {\it beaming models}, we find an optimum value of Lorentz
factor $\gamma$~=~8.5 for the radio quasars which are the beamed products
of the FRII parents; the BL\,Lac objects which are the beamed products of
the FRI parents show $\gamma$~=~15.0.  These values are not dissimilar
from those determined from VLBI observations of superluminal sources
\cite{ver95}.

The analysis accounts for major features of the source statistics from a
population definition which is physically meaningful. The increasingly
broad `evolution bump' in the source counts as survey frequency is
raised comes about naturally through the increasing intrusion of beamed
(flat-spectrum) objects. Despite the simplicity of assumptions, the
limited data defining the luminosity distribution and the small number of
parameters, the data are well described, as shown in Figure~\ref{count}.
Other tests are successful~\cite{jac99}, including source-count prediction
at different frequencies and the proportion of compact objects and
broad-line objects as a function of flux density.

%\section{Summary and considerations}

The success of the model demonstrates that essentially all radio AGN
detected in sky surveys above 1~mJy may be encompassed by the unification
hypothesis: {\it quasars and BL\,Lac objects are double-lobed radio
galaxies seen end-on}. At smaller flux densities, the population of AGN
declines and is replaced by the emergent population of starburst galaxies
(Figure~\ref{count}).

There are deficiencies. The model over-predicts counts at faint levels;
better estimates of the local luminosity function and of the
starburst-galaxy evolution (e.g. \cite{haa99}) should be incorporated.
Moreover it is now known that the redshift cutoff found for core-dominated
objects \cite{sha96} is a function of radio luminosity; the cutoff moves
to lower redshifts as radio luminosity decreases \cite{jac97}. At the
highest radio luminosities the space density profile with redshift
resembles the behaviour of the star-formation rate with epoch as
determined by Steidel~et~al.~\cite{ste99}. The simplistic `opera house'
models of figure \ref{lumdis} require modification accordingly. Finally
VLBA/VLBI surveys of core-dominated quasars show that there is a range in
jet speed \cite{kel99} with median values lower than those found here. How
these are related to the `apparent' jet speeds found here requires further
consideration. Such modifications are unlikely to destroy the basic tenet
of the analysis, that essentially all AGN found in radio surveys above
1~mJy can be described by unified (orientation-dependent) schemes. The
modifications will refine the definition of the AGN space-density profile,
and at such time the implications for associations between the AGN
phemomena and starburst activity may emerge with greater clarity.

%- dust-enshrouded quasars are unlikely to make a substantial contribution
%because the radio quasars are all seen.

%\section{References}

% References listed in alphabetical order ...

%
\vfill

{\small
\begin{moriondbib}

\bibitem{ter90} Terlevich E., Diaz A. \& Terlevich R.J., 1990, \mnras
{242} {271}
\bibitem{hec99} Heckman T., 1999, (Ringberg Conference) {\it
astro-ph/9903041}
\bibitem{mad96} Madau P., Ferguson H.C., Dickinson M.E., Giavalisco M.,
Steidel C.C. \& Fruchter A., 1996, \mnras {283} {1388}
\bibitem{wal98} Wall J.V., 1998, in {\it Observational Cosmology with the
New Radio Surveys} p.129, eds Bremer M.N. et al., Kluwer Academic
Publishers
\bibitem{boy98} Boyle B.J. \& Terlevich R.J., 1998, \mnras {293} {L49}
\bibitem{ste99} Steidel C.C., Adelberger K.L., Giavalisco M., Dickinson
M. \& Pettini M., 1999, {\it ApJ}, in press
\bibitem{wal97} Wall J.V. \& Jackson C.A., 1997, \mnras {290} {L17}
\bibitem{jac99} Jackson C.A. \& Wall J.V., 1999, \mnras {304} {160}
\bibitem{dun90} Dunlop J.S. \& Peacock J.A., 1990, \mnras {247} {19}
\bibitem{sch87} Scheuer P.A.G., 1987, in {\it Superluminal Radio Sources}
p.104, eds Zensus J.A. \& Pearson T.J., Cambridge University Press
\bibitem{bar89} Barthel P.D., 1989, \apj {336} {606}
\bibitem{bla78} Blandford R. \& Rees M.J., 1978, in {\it Pittsburgh
Conference on BL\,Lac Objects} p.328, eds Wolfe A.M. et al., University
of Pittsburgh
\bibitem{ant93} Antonucci R., 1993, {\it ARA\&A} {\bf 31}, 473
\bibitem{pea87} Pearson T.J. \& Zensus J.A., 1987, in {\it Superluminal
Radio Sources} p.1, eds Zensus J.A. \& Pearson T.J., Cambridge University
Press
\bibitem{fan74} Fanaroff B. \& Riley J.R., 1974, \mnras {167} {31P}
\bibitem{bro83} Browne I.W.A., 1983, \mnras {204} {L23}
\bibitem{mor91} Morris S.L., Stocke J.T., Gioia I.M., Schild R.E., Wolter
A., Maccacaro T. \& Della Ceca R., 1991, \apj {380} {49}
\bibitem{lai83} Laing R.A., Riley J.M. \& Longair M.S., 1983, \mnras
{204} {151}
\bibitem{wal80} Wall J.V., Pearson T.J. \& Longair M.S., 1980, \mnras
{193} {683}
\bibitem{sha96} Shaver P.A., Wall J.V., Kellermann K.I., Jackson C.A. \&
Hawkins M.R.S., 1996, \nat {384} {439}
\bibitem{cav99} Cavaliere A. \& Malquori D., 1999, \apj {516} {L9}
\bibitem{ver95} Vermeulen R.C., 1995, in {\it Quasars and AGN: High
Resolution Imaging} p.11385, eds Cohen M.H. \& Kellermann K.I., Publ.
Nat. Acad. Sci.
\bibitem{haa99} Haarsma D.B., Partridge R.B., Waddington I. \& Windhorst
R.A., 1999, in {\it Proceedings of the 19th Texas Symposium}, in press,
{\it astro-ph/9904036}
\bibitem{jac97} Jackson C.A., 1997, PhD Thesis, University of Cambridge
\bibitem{kel99} Kellermann K.I., Vermeulen R.C., Zensus J.A. \& Cohen
M.H., 1999, in {\it Proc 4th JIVE Symposium}, in press, eds Garett M.A.
et al., New Astronomy Reviews, Elsevier Science 
%\bibitem{urr95} Urry C.M. \& Padovani P., 1995, Publ. Astron. Soc. Pacif.  
%{107} {803}

\end{moriondbib}
}
\end{document}